# On the Nature of the El Niño/La Niña Events

David H. Douglass*, Drew R. Abrams, David M. Baranson
    Dept of Physics and Astronomy
    University of Rochester

B. David Clader
    Dept of Physics and Astronomy
    State University of NY at Geneseo

Classification:
    Major: Physical Sciences
    Minor: Geophysics

* Corresponding author. douglass@pas.rochester.edu

## Abstract

We propose a scenario that explains many of the Pacific Ocean climate phenomena that are called El Niño/ La Niña. This scenario requires an event, which we call a Super-Niño Event. It dominates other phenomena when it occurs. A template of this event has been constructed giving the time evolution, which is an alternating sequence of 'El Niños' and 'La Niñas'. The duration of the event is about 15 years unless some other event intervenes. Three such events can explain most of the El Niño/La Niña features that have been observed since 1968. We find that the various El Niño/La Niña features that have been observed fall into types, associated with the oscillation in the template, which can be classified by a "Periodic Table". The Earth is presently experiencing one of these events which started in the mid 1990's and will continue for another 4-5 years. This extrapolation into the future suggests that a minor El Niño will begin in mid 2001 and will reach a maximum about a year later. The conditions under which such a Super Niño-Event may occur are discussed.

    SST = Sea Surface Temperature
    SOI = Southern Oscillation Index
    Super Niño Event
    SST Template

# I. Introduction

There is considerable interest in the Pacific Ocean climate disturbances, El Niño (warm episodes) and La Niña (cool episodes), and their world-wide effects. The scientific measurements that seem to be related to these climate effects are the Sea Surface Temperature anomaly (SST) and the Southern Oscillation Index anomaly (SOI). SST is the deviation from a baseline of the mean temperature in a particular region of the Pacific Ocean and SOI is the pressure difference between two points, Darwin and Tahiti, also in the Pacific. (See Trenberth (1997) and Barnston et al. (1997). The geographic regions where these measurements are made are shown in fig.1 Measurements of SST and SOI for these regions are shown in figure 2A.

The definition of an El Niños/ La Niña event has been addressed by Trenberth (1997). He defines an event if the SST signal exceeds in magnitude 0.4 °C and lasts for at least 5 months; he has compiled a list of events satisfying these criteria. We have re-listed all his events since 1967 in table 1 along with other events that we will discuss.

Correlations between SST and SOI were shown by Bjerknes (1966) to be important and phenomena that involve both are referred to as El Niño/ Southern Oscillation (ENSO) effects. Barnston et al. (1997) showed that SOI measurements had the highest correlation at zero delay with the SST region designated Niño 3.4, which is the one we use.

These ENSO phenomena have been variously described as: 'having an interval of 3-7 years'; 'intervals with no La Niñas'; 'intervals with long El Niños being rare'; etc. One concludes that the characterization, the occurrence, and the magnitude of these events are not completely understood or well defined. The approach in this investigation is to study the largest events in the hope that these may be the key to greater understanding.

From the SST and SOI data in fig. 2A it is seen that, in fact, certain large amplitude events do stand out. The most recent 1997-98 event and before that the 1982-1983 event are two such examples. One might expect if an event were large enough, then it might dominate all other effects and that during this time interval one could learn about the intrinsic nature of the underlying phenomena. Perhaps it is not unlike where the study of the motion of the Earth during a large earthquake leads to knowledge of the normal modes of vibration and relaxation times of the Earth.

We have assumed that there is an underlying phenomenon associated with these large events and that when it occurs it is an independent event with an intrinsic behavior. We give it the name Super Niño (SN). Given this assumption, the task is to discover the nature of the event. From examination of the SST and SOI data we have found that three particular recent large events (the two mentioned above plus the 1972-1973 event) can be used to determine the SN event and its time evolution. It is determined quantitatively and is represented by a function, which we call the template.

The found SN template is an oscillating function of time with several maxima and minima, which are enumerated by indices [-2, -1, 0, +1, +2, +3]. See the schematic of fig 2B. The maxima are identified with El Niños and are given even index numbers; the largest maximum is the one associated with the term 'El Niño' as it is commonly used and has index 0. The minima are identified with La Niñas and have odd index numbers; the index +1 is the one usually associated with the term 'La Niña'.

Three such SN events are sufficient to explain most of the data. The observed El Niño/La Niña features in the data. are identified with the various maxima and minima of the SN template. In addition, we have created a classification scheme based upon the SN template and the sequence indices, which we call a 'Periodic Table'. This Periodic Table accommodates most of the events listed in table 1. The template can also be used to predict future values of SST and SOI. These results are explained and discussed below.

## II. The data

The Sea Surface Temperature (SST) data and Southern Oscillation Index (SOI) data are monthly values, which can be obtained on the web; see Garrett (2000).

These data are shown in fig 2A. The upper curve is for SST and the lower is SOI. The positive regions of SST and the negative regions of SOI are filled with red. These are associated in the literature with what are called El Niños. The other regions are filled with blue and are associated with what are called La Niñas. Visual inspection of these two data sets shows a strong correlation between the SST and SOI. We calculate a correlation r = -0.85. (This is close to the value -0.83 found by Barnston et al. [2] for the 1950-1996 interval.)

There are 3 large events labeled SN1 (72-73), SN2 (82-83), and SN3 (97-99) that we identify as Super Niño-Events (SN) [defined in table 2]. Inspection of any of the three SN in either data set shows an alternation of minimum and maximum values that is suggestive of a common pattern. The pattern that we have found is illustrated in the schematic shown in fig 2B. We show below that these 3 time segments are the only ones which show a large correlation with the template over the data range [1950 - 2000]. Thus the large SST peak at 1992 is not an SN event nor were any found from 1950 to 1969. Also there are various SST datasets prior to 1950, but none are of the integrity of the data we used.

## III. Analysis

### Stacking and Templates

The three selected large events are compared for similarity. They are described in table 2. The SST data for the 3 designated SN events were 'stacked'; i.e. plotted so that their maxima coincided (See fig 3A); this is sometimes called "compositing". The maxima, minima, and zero crossings of these 3 events are seen to be similar. This strongly suggests a common basic underlying intrinsic behavior, which can be expressed by a unique function, which we call a template

We constructed the SST template from the weighted average of the three
data sets in fig 3A and it is shown in fig 3B. A similar template can be constructed from
the SOI data using the SST times. A correlation r = 0.945, was found between the two
templates. Because of the high correlation, one can use either template. We chose SST,
and present the rest of the analysis with only SST. The beginning and end of the template
are not precisely known because of "noise." We show the "middle" 10 years.

## Comparisons Function and Forecast

An SST comparison curve, shown in fig4A, was generated for each SST SN event by
fitting the SST template to the maximum SST amplitude. The correlations of the
comparison curves with the SST data in the SN regions (time segments) are given in table
2. All three correlation coefficients exceed 0.9 A moving correlation of the SST template
against the entire range [1950 - 2000] of SST data was computed. The only values larger
than 0.5 are the three values in the table. We thus conclude that only the three regions
found can be described by the template.

It should be noted that the comparison curve for the SN3 event extends beyond the end of
the present data [July 2000]. This extension to 2004 constitutes a 'forecast' provided that
no other event intervenes. A magnification of fig 4A for the years 1994-2005 is shown in
fig 4B. Shown in fig 4B are the forecasts from Landsea (2000); CCA model and the
NCEP model [see Wang et al (2000)]. Landsea uses a model known as ENSO-CLIPER
(CLImatology PERsistence) as a basis for evaluating fourteen predictors. The Canonical
Correlation Analysis (CCA) model makes predictions based on patterns found in the
global SST. The National Center for Environmental Prediction (NCEP) has developed a
coupled ocean-atmosphere model.

# IV. Discussion and Conclusions

## The Super Niño-Event and the 'Periodic Table'.

A major result of this paper is that we have defined an event, which we call Super Niño
(SN), consisting of alternating positive and negative phases and is described by the
template shown in fig. 2B and fig. 3B. In the sequence, El Niños are indexed with even
integers and La Niñas with odd integers. The largest El Niño has index 0 and the largest
La Niña has index +1 The sequence corresponds to the observed El Niño/La Niña
features in the data. Three such SN's have occurred since 1968 The sequence 0, +1 is
often noted and has been discussed extensively. The extended sequence [-2, -1, 0, +1, +2,
+3] is noted for the first time here.

Another major result of this investigation is the construction of a 'Periodic Table' (See
fig.5), which classifies events according to the sequence indices [-2, -1, 0, +1, +2, +3].
The table has sequence number horizontally and time vertically. The time axis shows the
three different SN regions. We show that almost all of the El Niño / La Niña features observed since 1967 are in one of the SN event regions and each SN region shows a complete sequence.

The table lists 22 events. Events 1 to 18 (less 8 and 10) are explicitly listed by Trenberth (1997) as either an El Niño or as a La Niña, which we have further identified as belonging to a particular index. His suggestion that events 4 and 5 are part of one La Niña and that 6 and 7 are part of one El Niño agrees with our identification. Events 8 and 10 found by us are 'too weak' to have been detected by his criteria [magnitude greater than 0.4 °C for 6 months]. He further groups the 3 positive 'El Niños'( events 15, 16, and 17) together [This is called the region 'without La Niñas']. We find that events 15 and 16 lie outside of the range of the SN2 and SN3 template and thus must have a different explanation Event 17, however is identified by us as an $N_{-2}$ El Niño event associated with SN3. Event 18 is an $N_{-1}$ La Niña also associated with SN3.

Events 19 and 20 occurred after Trenberth's paper and are the recent El Niño and La Niña. Events 21 and 22 are 'forecasts' of this paper.

One sees that with the exception of events 15 and 16 that all observed events can be explained. There are only three Super Niño events hence only 3 type $N_0$ El Niños. All of the other El Niños in the list are related to these. For example the El Niño of 1986-88 is seen to be of type $N_2$ and is related to the $N_0$ El Niño of 1982-83 because both are phases of the SN2 event. Similar statements can be made concerning La Niñas.

## Regions (Time Segments) not Accounted for by Super Niño-Events

As noted above, the periods 1950 to 1969 and 1990 to 1994 showed some large features comparable to the features explained by SN events. So these events were not considered as SN events because of low correlation with the template. What is the explanation of these events? There are a number of possibilities.
1. The amplitude of the fluctuations that initiate a SN is small and the necessary threshold is not reached.
2. Fluctuations in amplitude are large enough but numerous. In this case an event, which starts, is soon interrupted by a later event.
3. Other geophysical effects dominate.
The calculations in the Appendix for the standard deviation of the annual term prior to 1968 gave much smaller values for the period 1950 to 1969. This favors explanation 1 for this region. However, an analysis the SST residuals given in table 2 for the SN periods shows that the SST amplitude for this period is larger than the amplitude of the residuals. This favors explanation 2. We prefer 2 to 1.

We spent considerable time studying the large 91-92 'El Niño'. It can not be fit into our scenario. It is possible that this is a SN event that got as far as the N-2 sequence when it was interrupted by the stronger SN3. Or explanation 3 may be invoked. We note that this particular 'El Niño' has been considered 'anomalous' by other investigators. See Trenberth and Hoar (1996).

## Hindcasts/Forecasts

As discussed above, the results indicate that the Earth is presently experiencing a SN event, which started in the mid 1990's and will continue for another 4-5 years unless interrupted. This extrapolation into the future suggests that a minor El Nino will begin in mid 2001 and will reach a maximum about a year later.

This 'prediction' is not the main result of our research efforts and we do not feel qualified to discuss this in terms that experts in this field use in predicting future climates. Nevertheless, it did come out of our research efforts so we show that result. We will let others evaluate what we have found using the technical terms of this field such as 'skill'. Our template is based upon only the record of the three SN events and the 'forecast' portion is based upon only two SN events. In order to obtain a more accurate template one would need many more records to reduce the error. Our subjective estimate of the error is 30-50%. So we would assign errors of this magnitude to the 'forecasts' of SST. According to our results, the Earth is presently in the late stages of a La Niña and is due to go into an $N_2$ El Niño phase early in 2001. Our 'forecast', however, is different from the others shown in fig 4B in that there should be a third negative dip before the zero crossing of the curve into the new El Nino region ($N_2$). This should happen in the next several months. We also note that the CCA model seems to suggest that such a dip will occur while the other two, Landsea and NCEP, do not. In addition, Landsea predicts a maximum as we do.

## Summary

The major results of this investigation are the discovery of the template describing what we call the Super Niño (SN) event. This template is used to construct a "periodic table" that allows a classification of most of the maxima and minima in the data. The template also can be used to make forecasts. In the appendix we suggest a possible model based upon parametric resonance.

## Addendum

This study was done during the summer of 2000. The analysis was done using data through July of 2000 when this paper was written. The forecast using the template start from that date. The delay between writing the paper and the final submission has allowed 11 SST data points to accumulate. A new figure [Fig. 6] has been added to the manuscript showing these 11 points, which can be compared to the forecast that we made. The data show a remarkable mirroring of the forecast through a maximum, a minimum, and a zero crossing. If the data follow the curve into the predicted small El Niño in late 2001 then one must consider that this scenario has some validity.

**Acknowledgements**. The SST template values are available upon request. This research was supported in part by the Rochester Area Community Foundation. Two of us [DMB and BDC} received a partial summer stipend under the University of Rochester NSF REU program.

# Appendix: Nature of the Super Niño-Event

The analysis in the body of the text is empirical and stands by itself with no attempt to relate the results to any model. In this appendix we discuss models and mechanisms. Whatever the process is, it must be the same for every event and explain both the beginning and end of the template.

**General Observations**
  1) The events are independent from each other.
  2) Since 1968, only three have occurred.
  3) There are periods of time that are not explained by SN events: prior to 1968 and from 1990-1994

As a function of time one can say:

  4) At a time prior to the time of the maximum of the template something begins exciting the ENSO system.
  5) At a time prior to the maximum, the amplitude is increasing.
  6) The time of the maximum is not the beginning of the SN event.
  7) After the maximum amplitude has occurred, the amplitude decreases with time.
  8) The event can be divided into two parts:
      (1) before the maximum: the excitation phase
      (2) after the maximum: the decay phase.

With this observation the maximum is thus the end of the excitation phase and the beginning of the decay phase.

  9) The duration of the event is greater than 15 years.

## Assumptions
We make the following assumptions about the SN in order to attempt an explanation of these observations.

  10) The state of the ENSO system in the absence of noise is zero before and after a Super Niño-Event.
  11) The ENSO system has normal modes, which have specific frequencies/periods. Even though one may not know what they are, they are fixed.
  12) That something excites one [or several] of these normal modes, which puts the ENSO system in a state in which the amplitude is larger than the 'noise'.
  13) The maximum is achieved when the excitation ceases [Or as suggested below, the excitation ceases when the maximum is reached].
  14) The amplitude of the normal modes will decay after excitation because of dissipative processes.

## Decay Phase

The decay phase is simpler to characterize than the excitation phase so we consider it first. Implicit is that we are concerned with the properties of the state of the system and not with the source of the excitation or the coupling. So the expectation that this part of the template is the same for all SN's is reasonable. Since the excitation is 'off' then one observes the 'impulse response' of the ENSO system. One of the simplest impulse responses is free decay at a frequency/period[s] of the normal mode[s]. From the template curve of fig. 3B one can infer

> D1. A normal mode[s] of period[s]  ~ 45-50 months.
> D2. A relaxation time  ~ 60-70 months.
> D3. A high frequency mode coupled to the 45-50 months mode.

The Fourier spectrum of the template has been computed and shows peaks consistent with these two periods.

## Excitation Phase

The excitation phase is harder to characterize. One needs:

> 15) A source of the excitation, which must contain frequencies related to the normal frequencies of the system that are observed to have been excited.
> 16) A mechanism to couple the source to the normal modes.
> 17) Sameness. Whatever accounts for 15) and 16) must be the same each time a SN event occurs if there is only one universal template.

Before continuing, it is important to note that there is a possible source of excitation present; namely, the strong forced annual oscillation, which is observed in the SST data. [Most discussions of SST are about the anomalies in SST data measured from a baseline that is defined by each month having its own average, which removes the annual and harmonic terms The base line values are different for each month; the largest values occur in Nov-Dec-Jan and the smallest in Apr-May-Jun. We will refer below to the plot of these values as the annual curve.]. Many investigators are probably unaware of this. This annual forcing term cannot satisfy requirement 15) directly because the frequency (period =12 months) is higher than the frequencies of the normal modes of this system (period = 45-50 months). However, one can couple the annual forcing to the normal modes if there are non-linearities in the system.

## Excitation Scenario

Here we consider a possible excitation scenario involving parametric resonance based upon an example from Landau and Lifshitz (1960). This scenario not only can explain the observations but also predicts properties not contained in the description of the template. Landau and Lifshitz consider the case of a damped resonant system with non-linearities which is being driven at a frequency $f_d$ that is about 3 times the resonant frequency $f_N$. The solution has the following properties:

LL1.   The response of the system is phase-locked to $f_d$ [period=12 months] but is at $f_d/3$ [period=36 months].
LL2.   The frequency $f_d/3$ must be close to, but greater than $f_N$.
LL3.   There is a minimum amplitude of the driving force [annual term] that is necessary in order to excite the system. In Landau and Lifshitz's words: " an initial 'push' is necessary in order to excite oscillations".
LL4.   There is a maximum amplitude beyond which excitation of the system is not possible.
LL5.   If the above conditions are not met, there are no solutions; the amplitude of the system at frequency $f_d/3$ is 0 even if the driving force at $f_d$ is not.

We compare the properties of the Super Niño-Event to the above list.

LL1'.   The template appears to show one cycle of oscillation at the required period of 36 months during the excitation phase.
LL2'.   The period during excitation [36 months] is less than the normal period [45-50 months] as required.
LL3'.   A threshold amplitude is required in order to initiate an event. Therefore, one would expect events to occur less frequently or not at all when the amplitude of the annual effect is small. We have calculated the standard deviation of the amplitude of the annual term [defined above] for different time intervals and find the results listed in table 3.
LL4'.   Maximum amplitude. The excitation once started would cause the amplitude to grow. If the excitation starts in Nov-Dec, then 36 months later in Nov-Dec the amplitude would be large. If this amplitude at that time exceeded the maximum value allowed then phase locking is lost The system is no longer coupled to the driving excitation even though it is still present. This accounts for the maximum occurring in those months. The system now enters the decay phase and the amplitude of the mode will now decay at the frequency $f_N$ of the normal mode.

One sees two fascinating results in the table:

(1) The amplitude of the standard deviation is lower during 1950-1968 where we have found no Super Niños than during 1968-1999 where we found three.
(2) The second result is even more interesting. The standard deviation is seen to have a large variation during the calendar year. The largest value occurs in Nov-Dec. This means that the minimum threshold requirement would most likely be satisfied in Nov-Dec. Thus, not only is the system phase locked in frequency to the annual driving term, it most likely to be excited in Nov-Dec.

To summarize this possible event scenario (ES):

ES1. The excitation of ENSO begins when a fluctuation in the amplitude of the annual driving force exceeds a certain threshold (Probably in Nov-Dec).

> ES2. The ENSO system is then phase-locked to the annual driving force and responds at a period of 36 months.
> ES2. The magnitude of amplitude builds until it reaches the maximum value allowed and coupling to the annual effect is not possible. This maximum may occur 36 months after ES1 (again in Nov-Dec). The observed maximum SST anomaly amplitude of 2.8 °C may be, in fact, this maximum allowed amplitude. [i.e. No larger El Niño events than have already been observed will occur!]
> ES4. The decoupled ENSO system then decays freely at period[s] 45-50 months with decay constant ~60 months [and exchanging energy with a coupled higher frequency mode].

The scenario in this appendix seems to not only to be consistent with the properties of the template; it predicts three new effects that are not a property of the template: minimum threshold amplitude, Nov-Dec phase, and a maximum amplitude, all of which are suggested by the data.

There is an extensive body of literature on ENSO models. Neelin et al. (1998) and Baraston et al. (1999) have summarized the various models of ENSO that have been considered. The models are grouped into two classes: dynamical and statistical. Although this template was found by statistical analysis the model presented in this appendix is definitely not in the statistical class. The papers of Zebiak and Cane (1987), Barnett et al. (1993), Kirtman et al. (1997), and Ji et al. (1996) are among those of the dynamic class. All of them are coupled models involving the atmosphere and the ocean. The Landau - Lifshitz (LL) model considered implies coupling because neither the ocean or the atmosphere has normal frequencies needed to explain the template. Some of these dynamical models have feedback but there is none here. Also one can say that this LL model is not a delayed oscillator model. In many of these models the driving force is unclear. In our scheme the driving force originates from the annual effect and the frequency is known and the mechanism is via parametric resonance. Our alternative scheme outlined above is not the same as any of these models.

This scenario is perhaps improbable in part or in total. However, even if wrong, it proves that the set of scenarios is not empty. There are undoubtedly many other excitation scenarios that one could invent but they would have to explain all of the details that we have noted. A challenge to this or any other plausible scenario is to understand the driving force and measured template evolution from the actual geophysical dynamics and climate physics.

# Legends

Fig. 1. Location Map of the pacific Ocean
　　Shows Niño region 3.4 where the SST measurements are made and the SOI stations at Tahiti and Darwin where the pressure difference measurements are made.

Fig. 2. Climate Index data and Super Niño-Event
　　(A) Sea Surface Temperature anomalies (SST) and
　　Southern Oscillation Index Anomalies vs. date.

　　(B) Schematic of Super Niño-Event. The region labeled
　　$N_0$ is what is called El Niño and the region labeled $N_1$ is what is called La Niña. The other labeled regions are features of the Super Niño-Event but have not been previously recognized.

Fig. 3. Stacking and Template Plots
　　(A) Stacking of the three SST Super Niño-Events. These events have their maximum at Nov 72; Jan 83; and Nov 97.
　　(B) Template derived from the stacking plots. The SST template from the 3 curves in (A).

Fig. 4. SST Data and Various Hindcasts/Forecasts
　　(A) Data, Comparison, Residuals. 1968 to 2004.The 3 Super Niño regions are indicated. Also shown is the not explained region 1990-1994.
　　(B)Same as A but for 1994-2004. Here the SST data is not averaged. Various forecasts beyond the present (July 2000) are also shown.

Fig. 5. El Niño and La Niña 'Periodic Table'

The El Niños and La Niñas since Sept 1968 are arranged into a "Periodic Table". A main result of this paper is that an event, which we call a Super Niño (SN) sometime, occurs consisting of alternating El Niños and La Niñas phases in a sequence [-2, -1,0, +1, +2, +3] given by the template described in this paper. Three such SN's have occurred during this interval of time accounting for most of the features observed in the data. El Niños are indexed with even integers with 0 corresponding to the largest and the one most often noted. La Niñas have odd indices with +1 being associated with the most often cited case. The sequence 0, +1 is often noted; the rest of the sequence is noted for the first time in this paper.
There are 22 such El Niño/ La Niña features addressed in this table. Features 1 to 18 (less 8 and 10) are explicitly identified by Trenberth [1] as either an El Niño or as a La Niña. His suggestion that 4 and 5 are part of one La Niña and that 6 and 7 are part of one El Niño is correct according to our scheme. The events 8 and 10 that have been found by us are 'too weak' to have been detected by Trenberth's criteria. We find that his positive features 15 and 16 do not fit our scheme; These are discussed in the text. Events 19 and 20 occurred after Trenberth's paper. Events 21 and 22 are forecasts of this paper

Fig. 6.  The delay between writing the paper and the final submission has allowed 11 SST data points to accumulate.  This figure has been added to the manuscript showing these 11 points, which can be compared to the forecast that we made.  The data show a remarkable mirroring of the forecast through a maximum, a minimum, and a zero crossing.  If the data follow the curve into the predicted small El Niño in late 2001 then one must consider that this scenario has some validity.

## Table Captions

Table 1.
	List of El Niño/ La Niña Events. Trenberth[1] gave an operational criterion of an event. An event occurs when the SST signal for region 3.4 exceeds 0.4 C and lasted for 5 months. Events 1 though 18 (less 8, 10) were generated by this criterion and are from his table 2. The La Niñas Events 8 and 10 found by us are too weak to have been identified by the Trenberth criterion. Trenberth couples the La Niña Events 4,5 and also the El Niño events 6,7 and 15,16,17. These events are discussed in the text. Events after 18 occurred after the last event in the list in the Trenberth paper and are discussed in the text also.

Table 2
	Properties of the Three Super Niño Regions
Table 3
	Standard deviation of Annual Term

| Table 1: List of El Niño/La Niña Events | | | |
|---|---|---|---|
| **Event** | **El Niño** | **La Niña** | **Comments** |
| 1 | Sep 68-Mar 70 | | |
| 2 | | Jul 70–Jan 72 | |
| 3 | Apr 72–Mar 73 | | |
| 4 | | Jun 73–Jun 74 | Grouped by Trenberth |
| 5 | | Sep 74–Apr 96 | |
| 6 | Aug 76-Jan 78 | | Grouped by Trenberth |
| 7 | Jul 77–Jan 78 | | |
| 8 | | Mar78-Jan79 | Missed by Trenberth Criterion |
| 9 | Oct 79–Apr 80 | | |
| 10 | | Jan81-Sep82 | Missed by Trenberth criterion |
| 11 | Apr 82–Jul 83 | | |
| 12 | | Sep 84–Jan 85 | |
| 13 | Aug 86–Feb 88 | | |
| 14 | | May 88-Jun 89 | |
| 15 | Mar 91–Jul 92 | | Grouped by Trenberth |
| 16 | Feb 93–Sep 93 | | |
| 17 | Jun 94–Mar 95 | | |
| 18 | | Sep 95– Mar 96 | |
| 19 | May97-Apr98 | | After Trenberth paper |
| 20 | | Jul98-(Mar00) | After Trenberth paper |
| 21 | (Jul01-Jan03) | | This paper |
| 22 | | (Mar03-Jan04) | This paper |

| Table 2: Three Super Niño Regions (time segments) | | | |
|---|---|---|---|
| Super Niño | SN1 | SN2 | SN3 |
| Date of Maximum | Nov 72 | Jan 83 | Nov 97 |
| Range of Super Niño Event | Apr 69 to May 79 | Jun 79 to Jul 89 | Apr 94 to Jul 00 |
| Correlation of template with data | 0.916 | 0.903 | 0.985 |
| Variance of data [($^o$C)*2] | 0.81 | 0.92 | 1.3 |
| Variance of residuals [($^o$C)*2] | 0.10 | 0.14 | 0.04 |
| Ratio of Variances (residuals/data) | 0.12 | 0.15 | 0.03 |

| Table 3: Standard Deviation of SST Annual Term | | | |
|---|---|---|---|
| **Time Interval** | **Low** | **Ave** | **High** |
| 1950- 1999 | 0.60 (May) | 0.87 | 1.21(Dec) |
| 1950-1968 | 0.50 (Apr) | 0.70 | 0.96(Nov) |
| 1969-1999 | 0.64 (May) | 0.96 | 1.38(Dec) |

**Fig. 1**

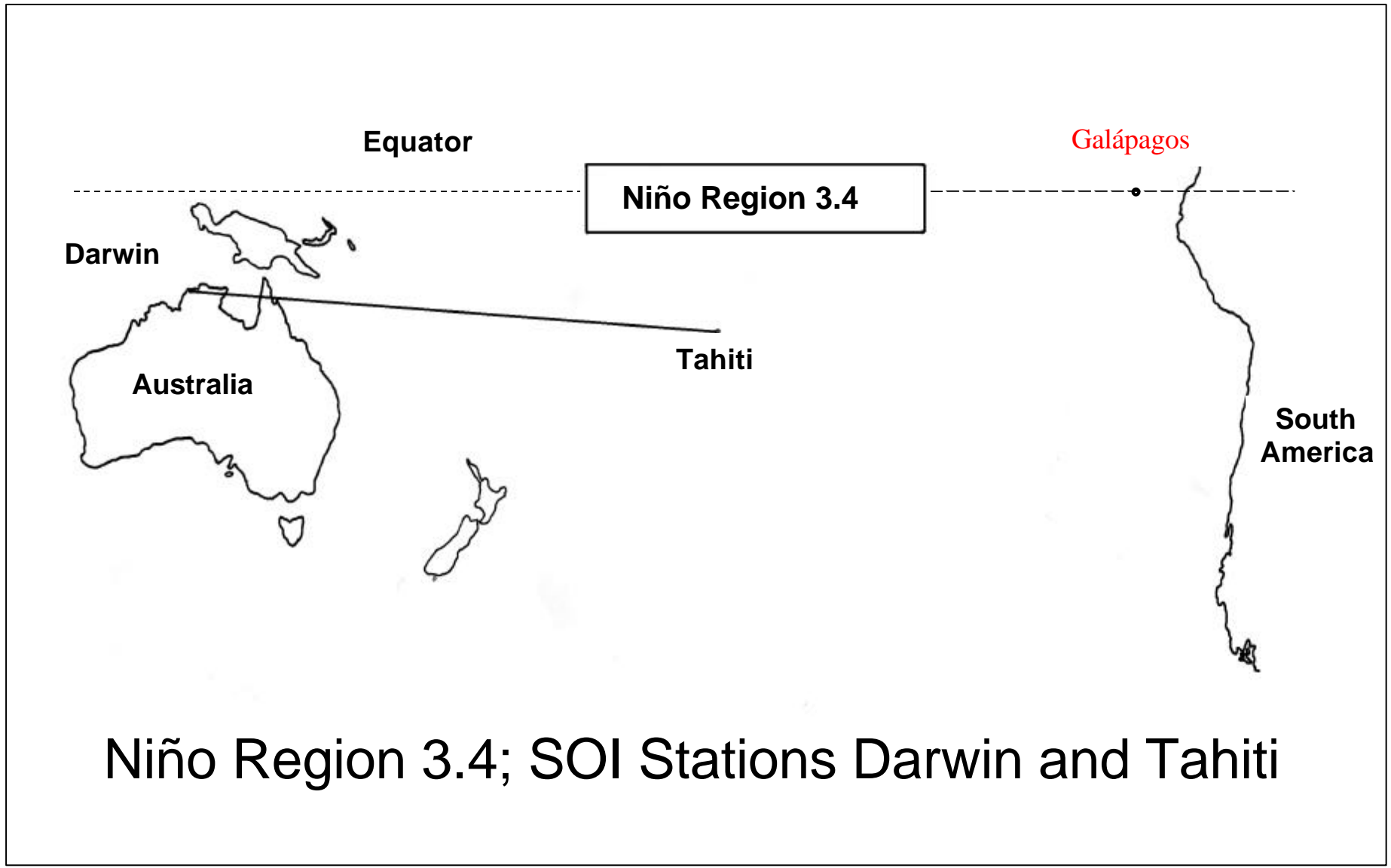

Niño Region 3.4; SOI Stations Darwin and Tahiti

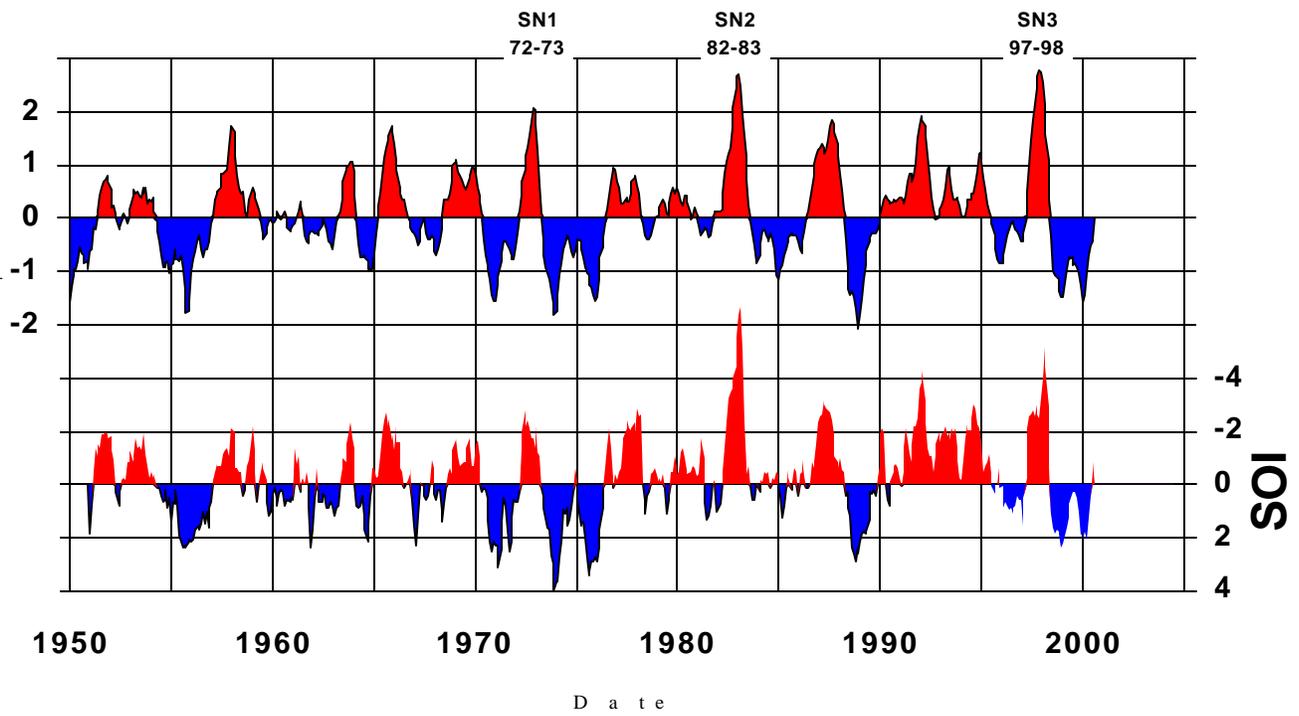

Fig. 2A SST and SOI

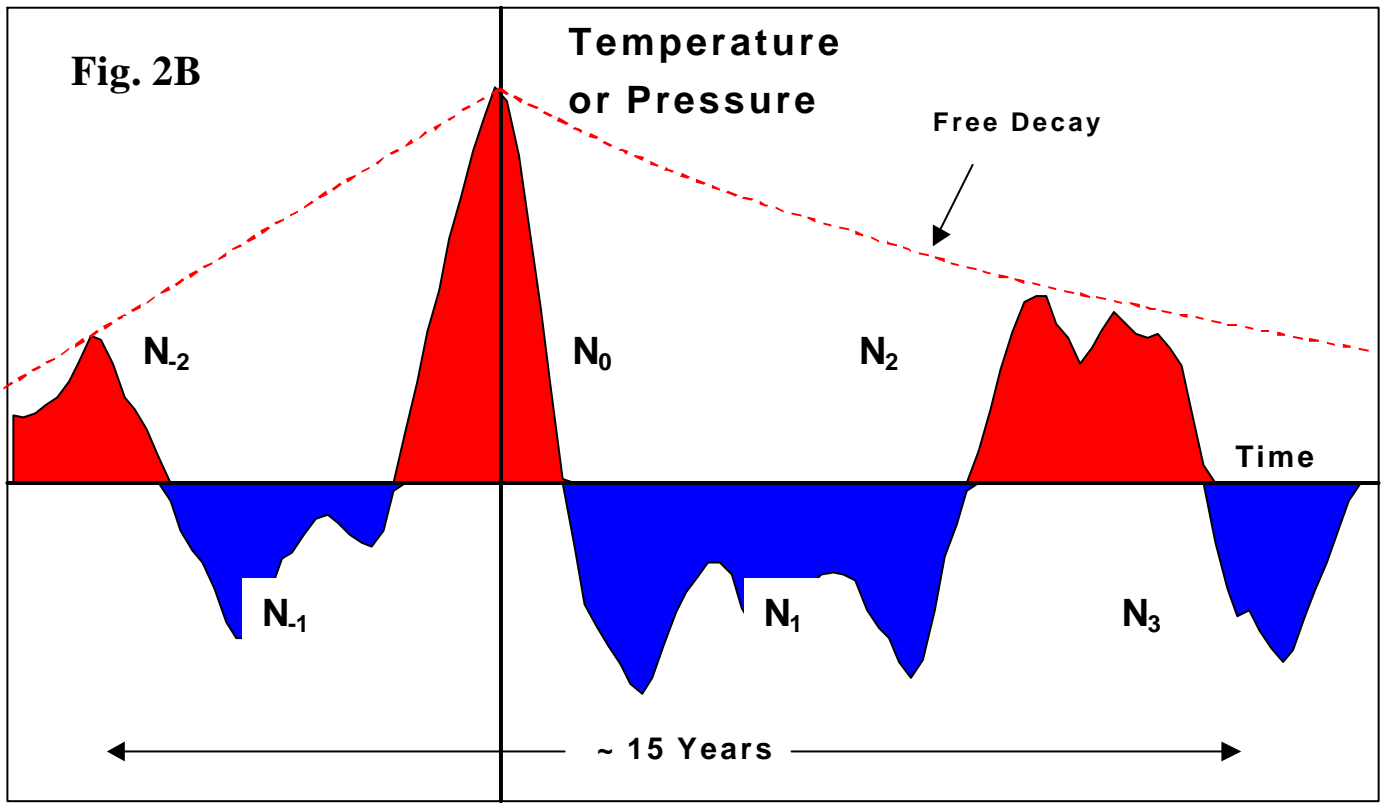

Fig. 2B

**Fig. 3A**

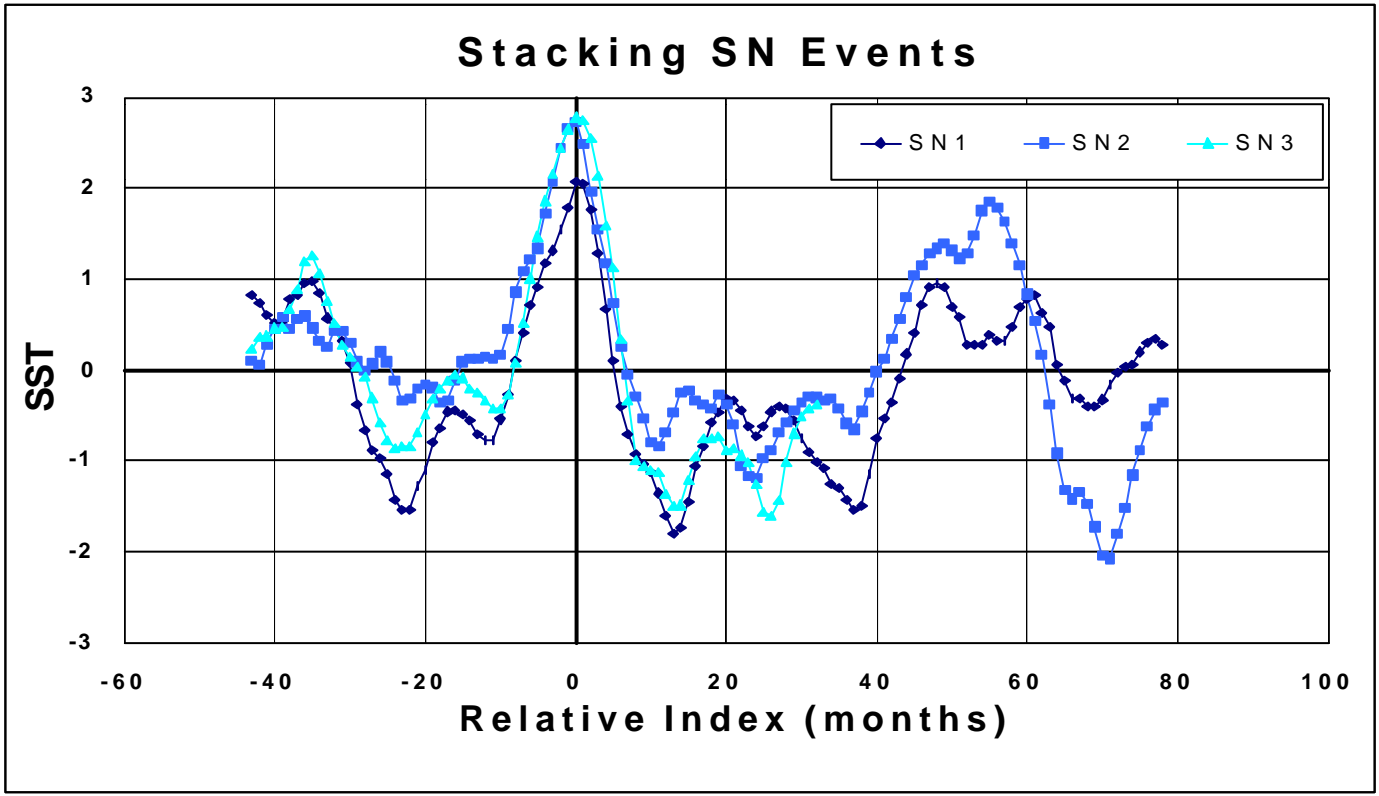

Fig. 3B

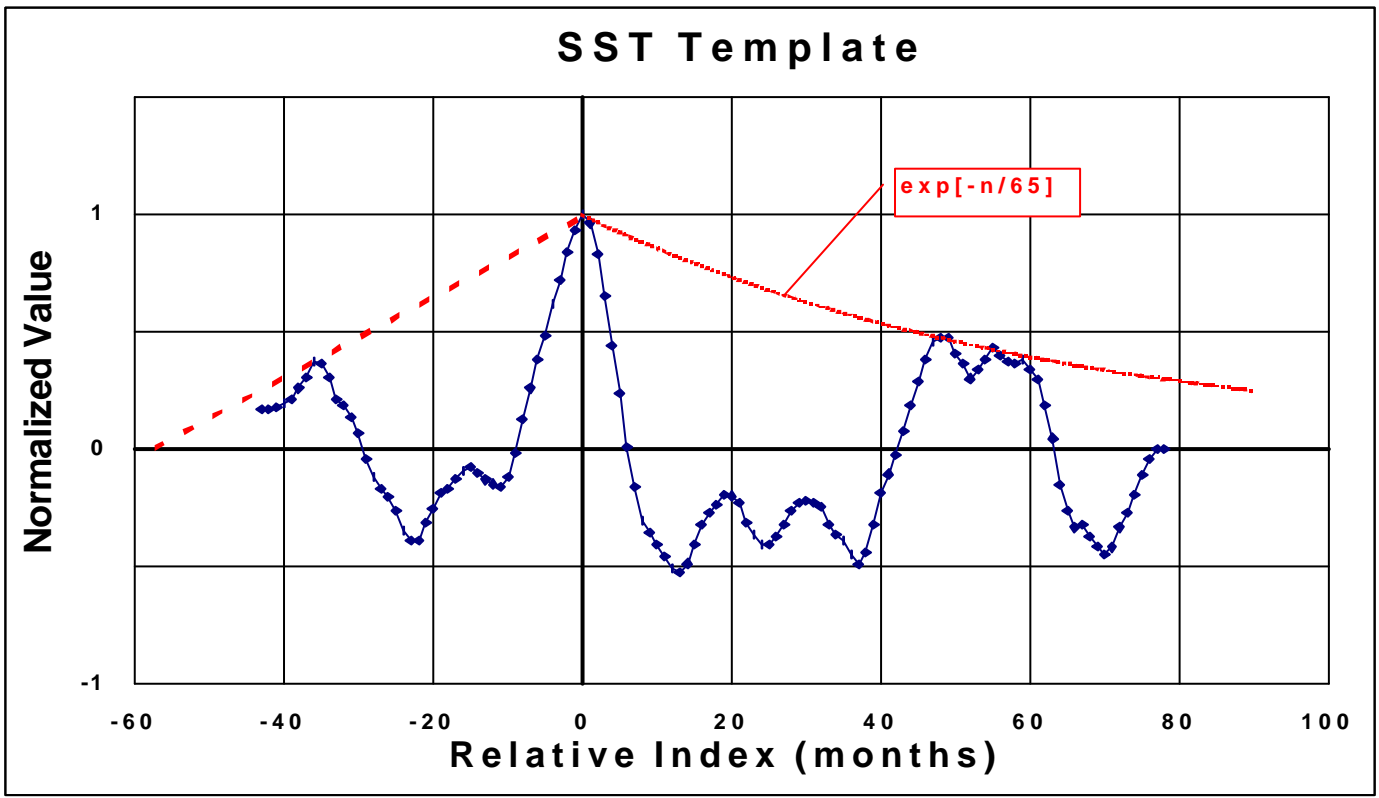

**Fig. 4A**

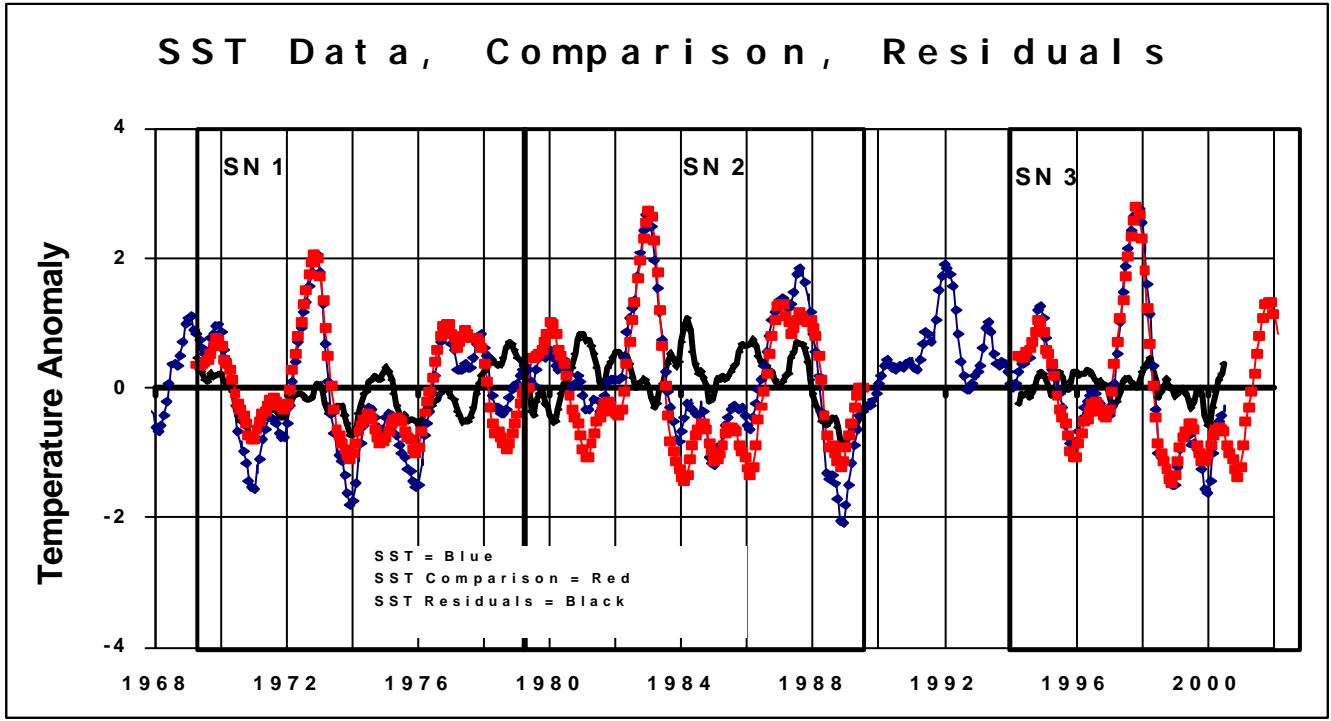

**Fig. 5A**

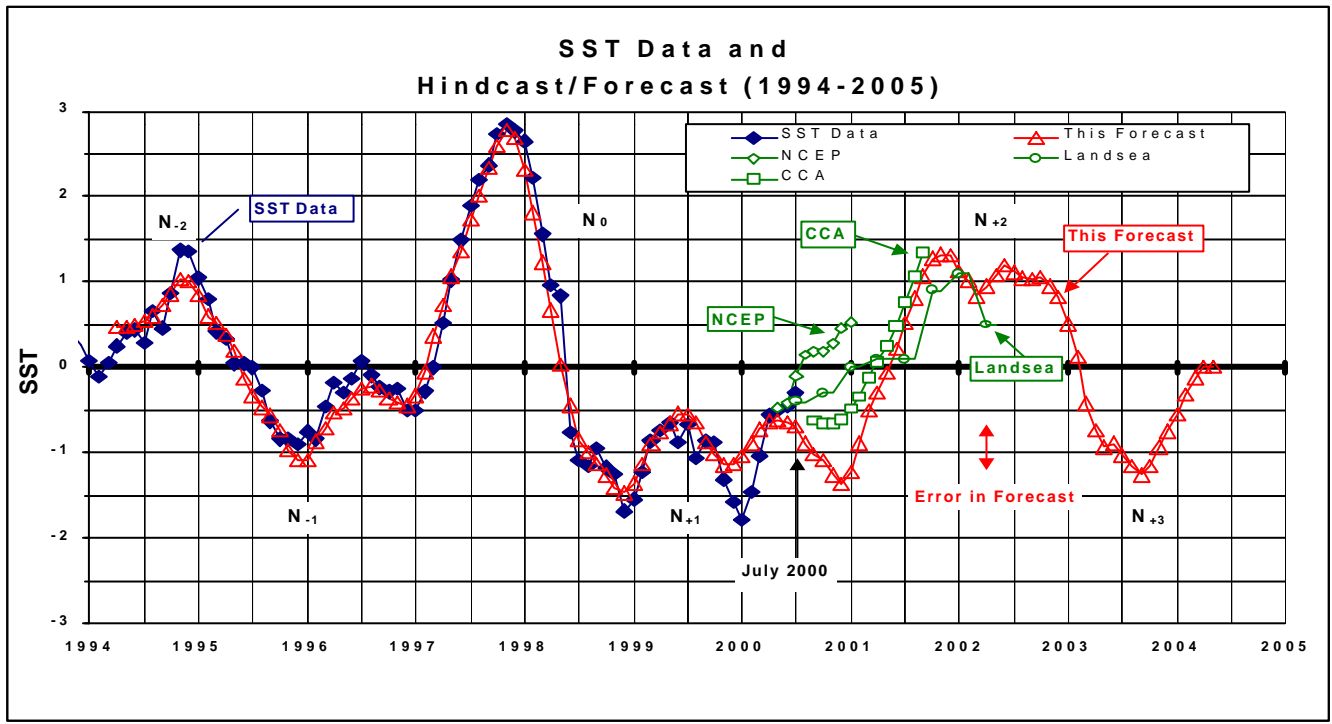

| Super Niño Region | Feature # | -2 El Niño | -1 La Niña | 0 El Niño | 1 La Niña | 2 El Niño | 3 La Niña |
|---|---|---|---|---|---|---|---|
| **SN1** | 1 | Sep 68 – Mar 70 | | | | | |
| | 2 | | Jul 70 – Jan 72 | | | | |
| | 3 | | | Apr 72 – Mar 73 | | | |
| | 4 | | | | Jun 73 – Jun 74 | | |
| | 5 | | | | Sep 74 – Apr96 | | |
| | 6 | | | | | Aug76 – Mar77 | |
| | 7 | | | | | Jul 77 – Jan78 | |
| | 8 | | | | | | (78-79) weak |
| **SN2** | 9 | Oct 79 – Apr 80 | | | | | |
| | 10 | | (81 – 82) weak | | | | |
| | 11 | | | Apr 82 – Jul 83 | | | |
| | 12 | | | | Sep 84 – Jun 85 | | |
| | 13 | | | | | Aug 86 – Feb 88 | |
| | 14 | | | | | | May 88 – Jun 89 |
| Two 'El Niños' | 15,16 | (Mar 91-Jul 92 and Feb 93 - Sep 93) listed by Trenberth [1] are outside of the range of SN template | | | | | |
| **SN3** forecast | 17 | Jun 94 – Mar 95 | | | | | |
| | 18 | | Sep 95 – Mar 96 | | | | |
| | 19 | | | May 97 – Apr 98 | | | |
| | 20 | | | | Jul 98 – (Mar 01) | | |
| | 21 | | | | | (Jul 01 – Jan 03) | |
| | 22 | | | | | | ( Mar 03 – Jan 04) |

Fig 5: El Niño and La Niña 'Periodic Table'

Not detected by the Trenberth [1] criteria.

**Fig. 6**

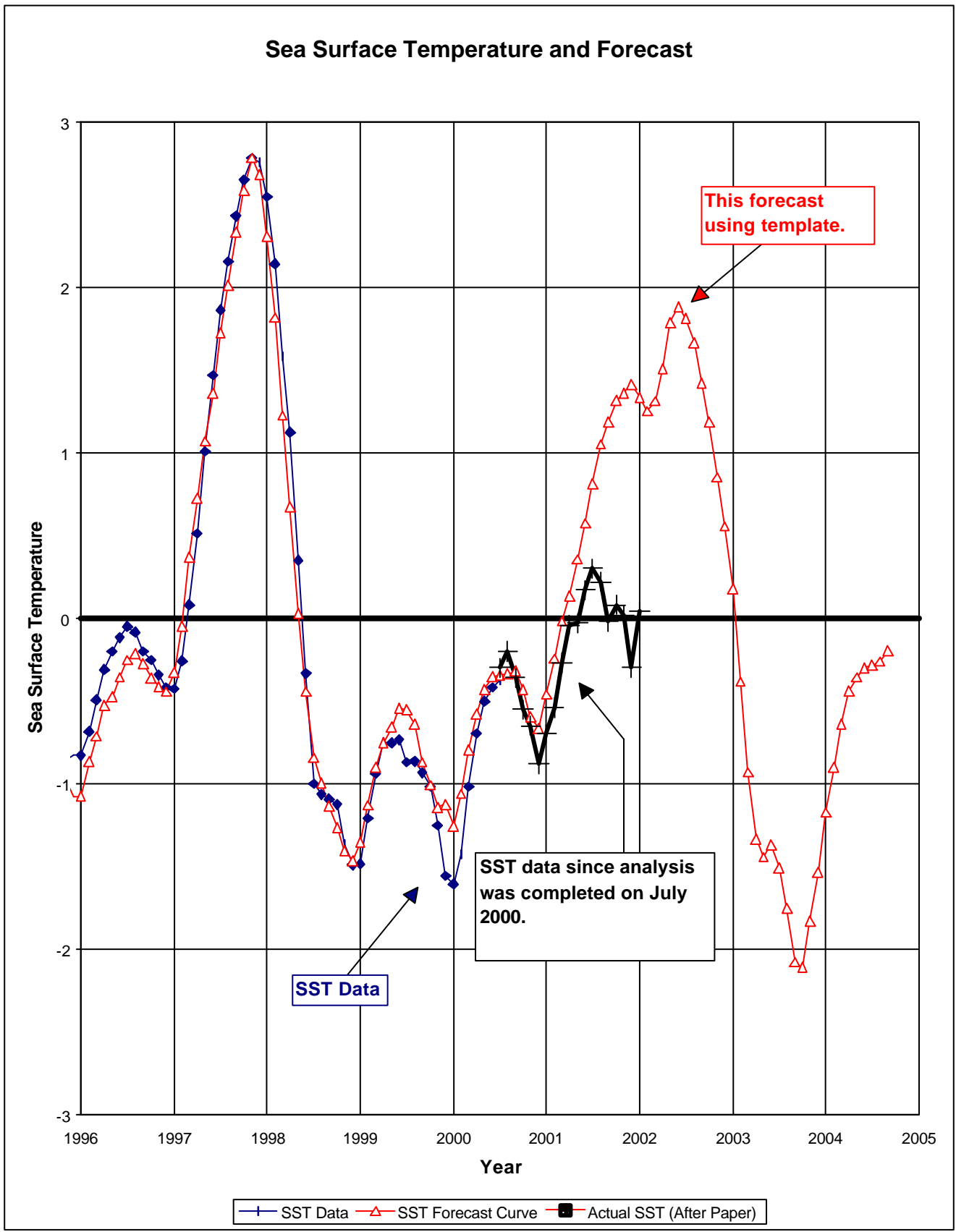